# Characterizing Efficient Referrals in Social Networks

Reut Apel
Technion IIT
Haifa, Israel
reutapel@campus.technion.ac.il

Elad Yom-Tov
Microsoft Research
Herzliya, Israel
eladyt@microsoft.com

Moshe Tennenholtz
Technion IIT
Haifa, Israel
moshet@ie.technion.ac.il

## ABSTRACT

Users of social networks often focus on specific areas of that network, leading to the well-known "filter bubble" effect. Connecting people to a new area of the network in a way that will cause them to become active in that area could help alleviate this effect and improve social welfare.

Here we present preliminary analysis of network referrals, that is, attempts by users to connect peers to other areas of the network. We classify these referrals by their efficiency, i.e., the likelihood that a referral will result in a user becoming active in the new area of the network. We show that by using features describing past experience of the referring author and the content of their messages we are able to predict whether referral will be effective, reaching an AUC of 0.87 for those users most experienced in writing efficient referrals. Our results represent a first step towards algorithmically constructing efficient referrals with the goal of mitigating the "filter bubble" effect pervasive in on line social networks.

## 1 INTRODUCTION

On line social networks are a popular medium for Internet users to obtain and disseminate information on topics of interest, including health, education, politics, and leisure activities. People are, however, usually exposed to only a small swath of the information available on these networks, because of the wealth of information available as well as the tendency of people to seek information which reinforces their viewpoints. The latter effect is known as selective exposure and, colloquially, as the "filter bubble" [3].

The ability to provide information outside people's normal areas of interest could increase user satisfaction, engagement, and social welfare [2]. When people search for news items, past research [3] found that the use of language congruent with that of the user can lead people to read opposing viewpoints. However, social networks differ from news searches, as currently information outside of the "filter bubble" is carried solely by users and administrators who know the relevant areas of knowledge and can refer people to them. Here we take the first step to an algorithmic solution to referrals, by asking which referrals are most likely to be associated with effective referral. While much work has been devoted to algorithmic aspects of information diffusion in populations and networks (see [1] and the references within) that work does not address the above fundamental challenge.

Specifically, we analyze the social media site Reddit (reddit.com). Reddit is organized around communities of shared interest. However, since there are many forums with similar topics, users do not always participate in the one most suitable for their needs, asking questions in one forum when the best answer for them can be found in another. Within each community (known as a subreddit) users open discussions or ask questions. These are **submissions**, each submission can attract responses and comments. A **referral** is when a comment to a question asked on Reddit invites the asker to look for the required information on another forum within Reddit. **Efficient referrals** (ERs) are those referrals where users who accept the invitation begin to participate in the recommended forum, in which they would not have participated prior to the referral.

The aim of this work is to characterize efficient invitations and distinguish them from all invitations. Our work follows two stages: First, a text-based classifier identified comments which contained referrals. Then, a classifier for distinguishing efficient from inefficient referrals was applied to a large corpus of comments.

## 2 DATA AND METHODS

Reddit is a platform for creating communities, and is built of thousands of active communities, known as 'subreddits'. Each subreddit is devoted to a different topic. Here we focused on a group of subreddits devoted to weight loss, we assumed that in this area of interest, there are many subreddits that are closed to each other in their topic, so people can influence others to participant in different subreddit with close topic.

Our initial goal was to filter Reddit comments as to whether or not they contain a referral. To this end, we collected 877 potential referrals and two people manually classified these comments as referral and non-referral based on the comment's content and by checking the comments to the submission. The agreement between annotators had a Cohen's Kappa of 0.78.

We used these labels to construct an automated classifier of comments for whether they contain a referral. Each comment was represented through the uni-, bi- and tri-grams of posts, weighted using TF-IDF, after eliminating stop words. We kept the 75% with the highest chi-square test values. We applied different learners and used leave-one-out to compare their AUC score.

Having identified comments which included a referral, we then proceeded to classify those referrals which were effective. To this end we extracted a total of 62,763 potential referrals comments (those including a link to other subreddits) and submissions from 44 of weight loss subreddits from December 2015 to February 2017 Of these, 16,700 included a referring link.

The comments were scored using the classifier with the highest AUC score from the first part, to find the probability of each of the comments to contain a referral. Comments with a probability greater than 0.9 were kept. Each of these were scored as to whether they were efficient by determining whether the author of the submission began posting in the recommended subreddit for the first time after the referral comment. Of 3,406 postings examined, 438 were efficient.

We built a model to identify ERs. The model was tested using 5-fold cross-validation, stratified by users. The following algorithms were tested: linear classifier with an L1 or L2 norm, multinomial and binomial naive Bayes models, linear SVM, Nearest centroid, and random forest. For lack of space we report the result of the best performing classifier. Each comment contains a referral was represented using the following attributes (in parenthesis, names of the attributes as referred to in Table 1). Note that all temporal attributes (e.g., past user interactions) were computed until just prior to the comment:

(1) Textual Attributes: Each word in the posting was represented by its TF-IDF value, enhanced by the word2vec representation of that word. Specifically, the weight of each word is: $\frac{1}{number\_of\_words} \sum word2vec * TfIdf$.
(2) Past user interactions: Number of comments and submissions in the original subreddit made by the submission and the referral authors ($N_{sub\_original}$ and $N_{com\_original}$, respectively), number of comments and submissions in the recommended subreddit made by the submission and the referral authors ($N_{sub\_reco}$ and $N_{com\_reco}$, respectively), number of referrals made by the referral author ($N_{ref-by-user}$), percent of ERs made by the referral author ($P_{eff}$).
(3) Temporal: Date and time of day of the submission and the referral ($Hour_{sub}$ and $Hour_{com}$, respectively), time from submission to referral ($Time_{between}$).
(4) Similarity of interactions: Similarity between the user's postings ($S_{users}$), similarity between the comment and the submission, and between the comment and the submission's title ($S_{com-sub}$ and $S_{com-title}$, respectively), similarity between the original and the recommended subreddits ($S_{subreddit}$).
(5) Referral attributes: Number of words (len), number of referrals in the comment ($N_{ref}$), number of responses to the submission ($N_{com-tree}$).

## 3 RESULTS

The model which best classified comments which contained a referral from those which did not, was a multinomial naive Bayes model, with an Area Under Curve (AUC) of 0.93.

We next compared different models to classify ERs. We assumed that past experience in writing ERs was of cardinal importance and therefore also split the data into three groups according to this attribute ($P_{eff}$).

Table 1 shows the results of a linear classifier, optimized using Stochastic Gradient Descent with an L2 norm. Our results suggest that classification of all data without stratification results in very poor performance. However, splitting the data by past efficiency allows us to reach good performance for users with a history ERs.

Beyond the cardinal importance of past experience in writing ERs, the best attributes for predicting ERs are past activity within the subreddit and the textual similarity of the comment to the submission. These results suggest that user's reputation (e.g., past subreddit activity) and using language similar to that of the asker lead to more predictably ERs.

Table 1: Best Classification results for the second part

| Data set | Features | AUC |
| --- | --- | --- |
| All data | $S_{subreddit}, N_{sub\_reco}, S_{com-sub}$ | 0.53 |
| $P_{eff} > 60\%$ | $N_{sub\_original}$, $Hour_{sub}$, $S_{com-sub}$, $N_{com\_reco}$, $Time_{between}, S_{subreddit}$ | **0.87** |
| $20\% < P_{eff} < 60\%$ | $N_{com-reco}, S_{users}$ | 0.60 |
| $20\% < P_{eff}$ | $N_{sub\_original}$, $N_{ref}$, len, $N_{com\_original}$, $N_{com\_reco}$, $N_{com}$, $Hour_{com}$, $N_{com-tree}$, $S_{subreddit}, Text\_Features$ | 0.55 |

## 4 CONCLUSIONS

On line social networks have been implicated in aggravation the natural tendency of people to read information congruent with their own viewpoints, an effect known colloquially as the "filter bubble". Being able to refer people to varied areas outside their usual areas of interest could potentially mitigate this effect. Here took the first step in understanding what effective referrals.

Our findings indicate that experience (as measured by past ability to efficiently refer people), past participation in discussions and the similarity of the referral to the question are all predictive of an ER. We note the importance of past experience in predicting future ERs, which was not captured in other attributes we used to model referrals, but allow a predictor which uses the referring author and the content of their messages, to predict whether a referral will be effective, reaching an AUC of 0.87 for users who are most experienced in writing ERs.

**Acknowledgements**: This project received funding from the European Research Council (ERC) under the European Union's Horizon 2020 research and innovation programme (grant agreement no. 740435).